\documentclass[11pt]{article}
\usepackage[margin = 2.5cm]{geometry}

\usepackage{hyperref}
\usepackage{graphicx}
\pdfoutput=1

\begin{document}
\title{Flow visualization inside a Ranque-Hilsch tube}
\author{Porta Zepeda David, Echeverr\'ia Arjonilla Carlos, \\ Stern Forgach Elizabeth Catalina 
\\
\\
\vspace{6pt} Taller de Hidrodin\'amica y Turbulencia, Facultad de Ciencias,  
\\ Universidad Nacional Aut\'onoma de M\'exico, Ciudad Universitaria,\\ MX 04510, M\'exico D.F.}
\maketitle

\begin{abstract}
In this fluid dynamics video we visualize the flow inside a Ranque-Hilsch tube either with baby powder or with water. 
\end{abstract}

\section{Introduction}
A Ranque-Hilsch tube is a mechanical device that, without any moving components, separates a stream of gas into a hot and a cold components. 
Air at $T_{in} = 17.5^{\circ} C $ and a high pressure of 5.72 atm. enters tangentially at a cross section of the tube Figure 1. 
In our configuration the hot stream exits at $T_{H} = 30.4^{\circ} C $ and the cold stream at $T_{C} = 14.9^{\circ} C$.
Our tube was constructed following the patent of Ranque [1].  The physical phenomenon has not been completely understood.

\begin{figure}[htbp]
	\centering
		\includegraphics[width=0.85\textwidth]{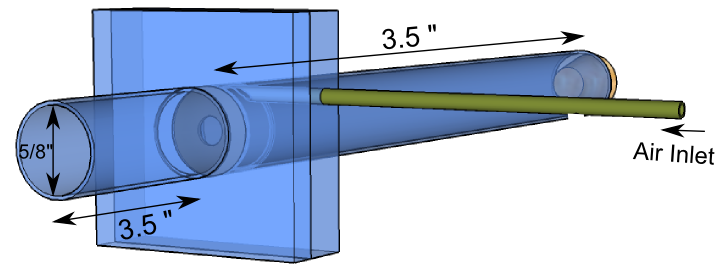}
	\caption{Ranque-Hilsch tube}
	\label{fig:Figura 1}
\end{figure}

The Ranque-Hilsch tubes (vortex tube) are now commercially used for low-temperature applications. 
In order to increase the efficiency, they have been proposed to replace the conventional expansion nozzle in refrigeration systems.

Before we seeded the flow we didn't know the existence of a swirling helicoidal motion, that can be observed in the slow motion videos (1200 fps, seeded with baby powder and water). We did expect the helicoidal mode.
To introduce the baby powder or the water, we made an atomizer as shown in Figure 2. 

\begin{figure}[htbp]
	\centering
		\includegraphics[width=0.90\textwidth]{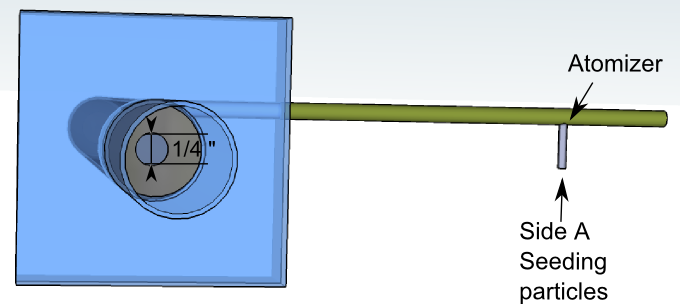}
		\caption{Side A seeding particles and cold side of the Ranque-Hilsch tube}
\end{figure}

\section{Acknowledgements}

Thanks to DGAPA-UNAM through Proyecto PAPIIT No.IN117712 "Propagaci\'on de ondas a trav\'es de superficies" and Professors Marcos Ley Koo and Andr\'es Porta.

\section{Reference}

[1]-US Patent No 1,952,281 from March, 1934. Ranque G.J. Method and Apparatus for Obtaining 
 \indent from Fluid under Pressure Two Currents  of Fluids at Different Temperatures. \\
 
\noindent [2]-Low-pressure vortex tubes. B Ahlborny, J Camirey and J U Kellerz 
 1996 J.  Phys. D: Appl. \indent Phys. 29 1469 \\ 
  
\noindent [3]-The Ranque effect. A F Gutsol 1997 Uspekhi Fizicheskikh Nauk, 
				Russian Academy of  Sciences \indent 1997 Phys.-Usp. 40 639

\end{document}